\documentclass[reprint,superscriptaddress,aps,prx,twocolumn,longbibliography]{revtex4-2}
\usepackage{makeidx}
\usepackage[utf8]{inputenc}
\usepackage{bm}
\usepackage{graphicx}
\usepackage{mathrsfs}
\usepackage{braket}
\usepackage[subrefformat=parens,labelformat=parens,caption=false]{subfig}
\usepackage{amsfonts}
\usepackage{float}    %
\usepackage{amssymb}
\usepackage{amsthm}
\usepackage{amsmath}
\usepackage{mleftright}
\usepackage{dcolumn}
\usepackage{color}
\usepackage[dvipsnames]{xcolor}
\usepackage{verbatim}
\usepackage{cancel}
\usepackage{multirow}
\usepackage[breaklinks,
            colorlinks,
            urlcolor=blue,
            linkcolor=blue,
            anchorcolor=blue,
            citecolor=blue]{hyperref}
           
\usepackage{soul}
\newcommand\norm[1]{\left\lVert#1\right\rVert}

\newtheorem{result}{Result}

\newcommand{\Eset}[1]{\underset{#1}{\mathbb{E}}}

\newcommand{\parens}[1]{\left(#1\right)}
\newcommand{\sparens}[1]{\left[ #1 \right]}
\newcommand{\bparens}[1]{\left\{ #1 \right\}}
\newcommand{\abs}[1]{\left\lvert #1 \right\rvert}
\DeclareMathOperator{\E}{\mathbb{E}}

\newcommand{\bigO}[1]{\mathcal{O}\mleft(#1\mright)}

\DeclareMathOperator{\Tr}{Tr}

\newcommand{\haar}{\mathcal{E}_{\text{Haar}}}
\newcommand{\U}{\mathcal{E}_{\text{U}}}
\newcommand{\gue}{\mathcal{E}_{\text{GUE}}}
\newcommand{\sff}[1]{\mathcal{R}_{#1}}

\begin{document}

\title{Entanglement governs early-time growth of randomness in projected ensembles}

\author{Thuwaragesh Jayachandran}
\email{tjayacha@caltech.edu}
\affiliation{California Institute of Technology, Pasadena, CA 91125, USA}
\author{Wai-Keong Mok}
\email{darielmok@caltech.edu}
\affiliation{California Institute of Technology, Pasadena, CA 91125, USA}

\begin{abstract}
Deep thermalization concerns the emergence of universal pure-state statistics in projected ensembles at late times, yet the mechanism governing the initial growth of randomness remains unclear. Here, we study the short-time dynamics of projected ensembles generated from initially unentangled states, quantifying their randomness using frame potentials. For arbitrary Hamiltonians, provided the initial bath state has full support in the measurement basis, we show that the frame potentials to cubic order in time are determined entirely by the subsystem purity, and hence by the bipartite entanglement generated between the unmeasured subsystem and its complement. The entanglement timescale therefore sets the initial timescale for the growth of local randomness, independently of the bath measurement basis. For unitarily invariant Hamiltonian ensembles, we further relate the projected-ensemble frame potential at order $k$ to the $4k$-point spectral form factor, or equivalently to the $2k$th frame potential of the global unitary dynamics, establishing a direct connection between local and global randomness. Our results identify entanglement as the mechanism governing the onset of randomness in projected ensembles and clarify how local randomness emerges from global quantum dynamics.
\end{abstract}
\maketitle

\section{Introduction}

Despite evolving unitarily, local observables of isolated quantum systems are generically described by equilibrium statistical mechanics at long times~\cite{nandkishor2015many,abanin2019many,deutsch1991quantum,srednicki1994chaos}. Quantum thermalization reconciles this apparent irreversibility with unitary dynamics: in the absence of additional conservation laws, the reduced density matrix of a local subsystem under chaotic global dynamics typically approaches a thermal Gibbs state~\cite{rigol2008thermalization,alessio2016quantum,borgonovi2016quantum,mori2018thermalization,ueda2020quantum,gogolin2016equilibration}.

Recent work has revealed finer structure beyond the coarse-grained description provided by the reduced density matrix, by studying the \textit{projected ensemble} on a local subsystem, which is obtained by projective measurements on its complement~\cite{cotler2023emergent,choi2023preparing}. The projected ensemble can be thought of as an unraveling of the reduced density matrix. For generic quantum many-body chaotic systems at late times, the \textit{projected ensemble} has been proposed to approach the (generalized) \textit{Scrooge} ensemble~\cite{goldstein2006distribution,goldstein2016universal,mark2024maximum}. The Scrooge ensemble is a distribution of pure states that realizes an arbitrary fixed density matrix (for example, a thermal Gibbs state) while being `maximally stingy' with the classical information extractable through measurements~\cite{josza1994lower,mcginley2025scrooge,mok2026nature}. This phenomenon, known as \textit{deep thermalization}, extends quantum thermalization to the higher-order fluctuations of the underlying pure-state distribution, and uncovers new universal behavior in quantum many-body systems~\cite{cotler2023emergent,choi2023preparing,ippoliti2022solvable,ho2022exact,wilming2022high,claeys2022emergent,shrotriya2023nonlocality,liu2024deep,bhore2023deep,lucas2023generalized,mark2024maximum,chan2024projected,chang2025deep,varikuti2024unraveling,mok2025optimal,liu2025coherence,zhang2025holographic,yan2025characterizing,chakraborty2025fast,bejan2025matchgate,manna2025projected,vairogs2025extracting,loio2025quantum,goldstein2006distribution,goldstein2016universal,varikuti2025deep,mok2026nature,liu2026conditional,feng2026quantum}.

At infinite effective temperature, the Scrooge ensemble reduces to the Haar ensemble, a uniform distribution of random pure states. More specifically, the projected ensemble is expected to form approximate state \textit{$k$-designs}, whose first $k$ moments are statistically indistinguishable from the Haar distribution~\cite{dankert2009exact,gross2007evenly,ambainis2007quantum}.

Previous work has largely focused on the late-time equilibrium behavior of projected ensembles~\cite{cotler2023emergent,mark2024maximum,mok2026nature}, as well as how projected ensembles dynamically approach $k$-designs in various models of quantum dynamics~\cite{ippoliti2022solvable,chan2024projected,varikuti2024unraveling,ghosh2025design,ippoliti2023dynamical,liu2026emergence}. A general framework describing the nonequilibrium dynamics of projected ensembles was recently developed by Anza and Hahn~\cite{anza2026nonequilibrium}. Here, we provide a complementary perspective by instead studying the early-time regime, in which the projected ensemble remains far from the Haar ensemble or, more generally, the Scrooge ensemble. We ask: \textit{what governs the initial growth of quantum randomness in a projected ensemble?} Restricting to the infinite-temperature case, we quantify this randomness using frame potentials, which probe successive statistical moments of the projected ensemble~\cite{roberts2017chaos}.

Our main result establishes a quantitative connection between the randomness of the projected ensemble and the bipartite entanglement generated between the unmeasured subsystem and its complement. We show that, for every moment order $k$ and any Hamiltonian, provided the initial bath state has full support in the measurement basis, the early-time frame potential is controlled by the second R\'enyi entanglement entropy up to cubic order in time (Result~\ref{res:1}). Because entanglement is basis independent, the influence of the measurement basis---which is crucial at late times---can first appear at quartic order and higher. Relating this behavior to the \textit{universal entanglement timescale}~\cite{yang2018entanglement,cresswell2018universal}, we find that the timescale governing the initial growth of randomness is the entanglement timescale set by the interaction between the subsystems (Result~\ref{res:2}).

Having identified entanglement as the generic mechanism, we next ask how this initial growth reflects the spectral properties of the Hamiltonian. For unitarily invariant Hamiltonian ensembles, such as the Gaussian Unitary Ensemble (GUE), we first show that the initial growth depends only on the spectral variance (Result~\ref{res:3}). We then relate the $k$th frame potential of the projected ensemble to the $4k$-point spectral form factor, or equivalently to the $2k$th frame potential of the global unitary ensemble~\cite{cotler2017chaos} (Result~\ref{res:4}). This relation reveals a ``$2k \to k$'' correspondence: randomness of the global unitary ensemble at order $2k$ controls the initial growth of randomness in the projected ensemble at order $k$. The correspondence complements previous theorems relating approximate unitary $2k$-designs in the global dynamics to approximate state $k$-designs in the projected ensemble~\cite{ghosh2025design,mok2026nature}. %

Our results establish a hierarchy in the onset of quantum randomness. Under the full-support assumption, entanglement universally governs the leading behavior at short times, while the effect of the measurement basis can appear only at higher orders. For unitarily invariant ensembles, the initial growth depends only on the spectral variance and therefore does not by itself diagnose quantum chaos. Our work thus provides a starting point for understanding when finer signatures of deep thermalization and chaotic dynamics first appear.

This paper is organized as follows. We introduce the basic notions of projected ensembles and frame potentials in Sec.~\ref{sec:projected_ensemble}. In Sec.~\ref{sec:general}, we derive a general formula for the early-time frame potential under arbitrary Hamiltonian evolution. In Sec.~\ref{sec:unitarily_invariant}, we specialize to unitarily invariant Hamiltonian ensembles and relate the frame potential of the projected ensemble to spectral form factors. We conclude in Sec.~\ref{sec:conclusion} with a discussion of our results and future directions.

\section{Projected ensemble and quantum randomness}\label{sec:projected_ensemble}

Consider a bipartite quantum system $AB$ initialized in a pure state $\ket{\Psi_0}$, with subsystem dimensions $D_A$ and $D_B$. A time-independent Hamiltonian $H$ generates the state $\ket{\Psi_t}=e^{-iHt}\ket{\Psi_0}$ at time $t$, where we set $\hbar=1$. We then measure subsystem $B$ in a fixed orthonormal basis $\{\ket{z}\}$, with $z\in\{1,\ldots,D_B\}$. Throughout this work, we take this to be the computational basis for concreteness, although our early-time results apply to any basis in which the initial bath state has full support. The Born probability of measuring outcome $z$ is
\begin{equation}
    \label{eq:pz}
    p(z) = \braket{\tilde{\psi}_z|\tilde{\psi}_z},
\end{equation}
where
\begin{equation}
    \label{eq:psi_z}
    \ket{\tilde{\psi}_z} = (I_A \otimes \bra{z}_B)\ket{\Psi_t}.
\end{equation}
Conditioned on the measurement outcome $z$, the quantum state on subsystem $A$ is described by the normalized projected state
\begin{equation}
    \ket{\psi_z} = \frac{\ket{\tilde{\psi}_z}}{\sqrt{p(z)}} = \frac{\ket{\tilde{\psi}_z}}{\sqrt{\braket{\tilde{\psi}_z|\tilde{\psi}_z}}}.
\end{equation}
These conditional states and their probabilities define the \emph{projected ensemble} $\mathcal{E}=\{p(z);\ket{\psi_z}\}$ generated by $\ket{\Psi_t}$.

For a chaotic Hamiltonian $H$ at late times $t$, and an initial state $\ket{\Psi_0}$ at infinite temperature with respect to $H$, deep thermalization posits that the projected ensemble is generically described by the Haar ensemble $\haar$, in the thermodynamic limit $D_B \to \infty$ for fixed $D_A$~\cite{cotler2023emergent,choi2023preparing}. The Haar ensemble, $\haar$, is a uniform distribution of quantum states on the Hilbert space of subsystem $A$. Here, the measured subsystem $B$ acts as the environment (or bath) for subsystem $A$, the system of interest. %
Unlike tracing out the environment, measuring $B$ retains the outcome-resolved pure-state distribution on $A$ and thereby probes quantum randomness beyond the reduced density matrix. Intuitively, one can regard the projected ensemble as an unraveling of the reduced density matrix.

To quantify the randomness of the projected ensemble $\mathcal{E}$, it is useful to compare its $k$th moment operator (for $k \in \mathbb{N}$),
\begin{equation}
    \label{eq:rhok}
    \rho_{\mathcal{E}}^{(k)} = \Eset{\ket{\psi} \sim \mathcal{E}} \sparens{\parens{\ket{\psi}\bra{\psi}}^{\otimes k}} = \sum_{z=1}^{D_B} p(z) \parens{\ket{\psi_z}\bra{\psi_z}}^{\otimes k},
\end{equation}
with the $k$th moment operator of the Haar ensemble $\haar$,
\begin{equation}
    \rho_{\text{Haar}}^{(k)} = \Eset{\ket{\phi} \sim \haar} \sparens{\parens{\ket{\phi}\bra{\phi}}^{\otimes k}}.
\end{equation}
We use the normalized Hilbert--Schmidt (Frobenius) distance between the $k$th moments of $\mathcal{E}$ and $\haar$ on $A$ to measure the degree of randomness:
\begin{equation}
    \label{eq:HSdistance_def}
    \Delta^{(k)}(t) = \frac{\norm{\rho_{\mathcal{E}}^{(k)} - \rho_{\text{Haar}}^{(k)}}_2}{\norm{\rho_{\text{Haar}}^{(k)}}_2},
\end{equation}
where $\norm{\cdot}_2$ is the Schatten 2-norm. The distance can be rewritten as~\cite{mok2025optimal}
\begin{equation}
    \label{eq:HSdistance}
    \Delta^{(k)}(t) = \parens{\frac{F^{(k)}(t)}{F^{(k)}_\text{Haar}}-1}^{1/2},
\end{equation}
where
\begin{equation}
\label{eq:framepot_trace}
    F^{(k)}(t) \equiv \Tr\sparens{{\parens{\rho_{\mathcal{E}}^{(k)}}^2}}
\end{equation}
is the $k$th frame potential of $\mathcal{E}$ (i.e., the purity of $\rho_{\mathcal{E}}^{(k)}$) and
\begin{equation}
    F_{\text{Haar}}^{(k)} = \Tr\sparens{{\parens{\rho_{\text{Haar}}^{(k)}}^2}} = \binom{D_A + k - 1}{k}^{-1}
\end{equation}
is the $k$th frame potential of $\haar$~\cite{cotler2017chaos}. This is also the inverse dimension of the $k$-fold symmetric subspace of subsystem $A$~\cite{harrow2013church}. We therefore focus on the frame potential
\begin{align}
\label{eq:framepot}
F^{(k)}(t) 
= \sum_{z_1, z_2} p(z_1) p(z_2)
\abs{\braket{\psi_{z_1}|\psi_{z_2}}}^{2k},
\end{align}
with lower value of $F^{(k)}$ indicating more quantum randomness.

\subsection{Dependence on measurement basis}

It is important to emphasize the key qualitative difference between $k = 1$ and $k > 1$. The moment operator of $\mathcal{E}$ for $k = 1$ is simply
\begin{equation}
    \rho_\mathcal{E}^{(1)} = \Tr_B \parens{\ket{\Psi_t}\bra{\Psi_t}},
    \label{eq:rho1}
\end{equation}
the reduced density matrix on subsystem $A$, while $\rho_{\mathcal{E}}^{(k)} \neq \parens{\rho_{\mathcal{E}}^{{(1)}}}^{\otimes k}$ in general for $k > 1$. Thus $F^{(1)}=\Tr\rho_A^2$ is the subsystem purity and $S_2(A)=-\log F^{(1)}$ is the second-R\'enyi entanglement entropy; both are independent of the measurement basis on $B$. Higher frame potentials probe basis-dependent effects in the higher-order statistical moments of the projected ensemble. As we will now show, provided the initial bath state has full support in the measurement basis, the early-time behavior of $F^{(k)}(t)$ is independent of the measurement basis for all $k$ up to cubic order in time.

\section{General formula for early-time frame potential}\label{sec:general}

\begin{figure*}[t!]
    \centering
    \includegraphics[scale =0.5]{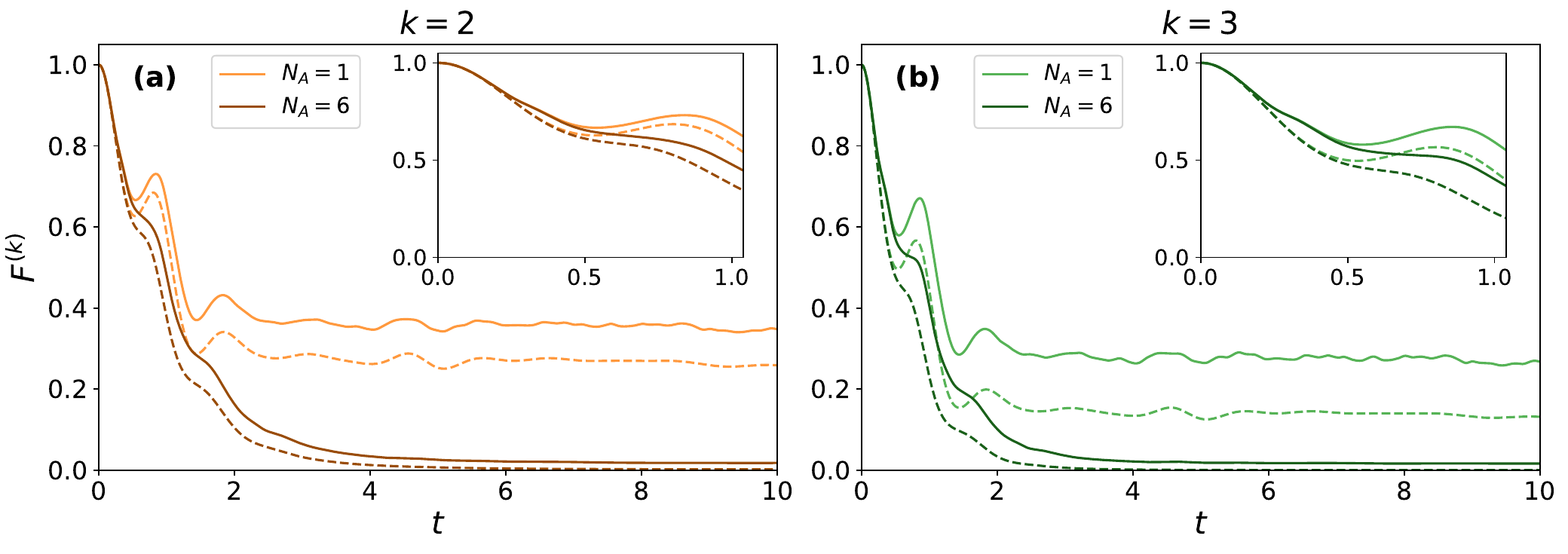}
    \caption{Numerical $F^{(k)}(t)$ (solid lines) and $(F^{(1)}(t))^k$ (dashed lines) for (a) $k=2$ and (b) $k=3$ in the 1D mixed-field Ising model with bath size $N_B=6$. %
    The inset shows a magnified view for $t\in[0,1]$, where $(F^{(1)}(t))^k$ agrees well with $F^{(k)}(t)$ up to $Jt\approx 0.2$.}
    \label{fig:mfim_result}
\end{figure*}

We now derive general expressions for the $k$th frame potential under an arbitrary time-independent Hamiltonian $H$, subject to assumptions on the initial pure state $\ket{\Psi_0}$ specified below.

Equation~\eqref{eq:framepot} is the expectation value of $\abs{\braket{\psi_{z_1}|\psi_{z_2}}}^{2k}$ over the joint distribution $\mathcal{P}$ defined by $p(z_1,z_2)=p(z_1)p(z_2)$. Define
\begin{equation}
    \label{eq:delta}
    \Delta_{z_1, z_2} (t)\equiv 1-  \abs{\braket{\psi_{z_1}|\psi_{z_2}}}^{2}.
\end{equation}
The $k$th frame potential can then be written as (suppressing the $t$, $z_1$, and $z_2$ dependence of $\Delta_{z_1,z_2}(t)$)
\begin{align}
    \label{eq:framepot_as_expectation}
    F^{(k)}(t) &=  \sum_{z_1, z_2} p(z_1) p(z_2)
\parens{1-\Delta_{z_1,z_2}}^{k}\\
    &=\E_{\mathcal{P}} \parens{1-\Delta}^{k}\\
    &= \sum_{l=0}^k \binom{k}{l} (-1)^{l} \E_{\mathcal{P}}\parens{\Delta^l}\\
    \label{eq:framepot_as_expectation_final}
    &= \sum_{l=0}^k \binom{k}{l} (-1)^{l} M_l,
\end{align}
where $M_l\equiv\E_{\mathcal{P}}\parens{\Delta^l}$. Trivially, $M_0=1$. For $l=1$,
\begin{equation}
    \label{eq:delta_1}
    M_1(t) = 1-\E_{\mathcal{P}}\parens{\abs{\braket{\psi_{z_1}|\psi_{z_2}}}^{2}} =1-F^{(1)}(t).
\end{equation}
Thus, Eq.~\eqref{eq:framepot_as_expectation_final} becomes
\begin{equation}
\label{eq:fk_f1_general}
     F^{(k)}(t) = 1 -k(1-F^{(1)}(t)) +\sum_{l=2}^k \binom{k}{l} (-1)^{l} M_l(t).
\end{equation}

Equation~\eqref{eq:fk_f1_general} is valid at all times. To obtain explicit early-time results, we assume that the initial state is a product state,
\begin{equation}
    \label{eq:Psi_0}
    \ket{\Psi_0} = \ket{\phi_A} \otimes \ket{\phi_B},
\end{equation}
where $\ket{\phi_A}$ and $\ket{\phi_B}$ are pure states on $A$ and $B$, respectively, so the subsystems are initially unentangled. We also assume that $\ket{\phi_B}$ has full support in the measurement basis:
\begin{equation}
    \label{eq:psi_0_assumption}
    \braket{z|\phi_B}\ne 0 \quad \forall z \in \{1,\ldots,D_B\}.
\end{equation}
Under these assumptions, the projected ensemble at initial time $t=0$ becomes
\begin{equation}
    \mathcal{E}_{0} = \{p(z,t=0); \ket{\psi_z, t=0}\} = \{\abs{\braket{z|\phi_B}}^2; \ket{\phi_A}\},
\end{equation}
For any product initial state, every outcome of nonzero probability produces $\ket{\phi_A}$, and hence $F^{(k)}(0)=1$. %

We decompose the normalized projected states $\ket{\psi_z(t)}$ into components along the initial state $\ket{\phi_A}$ and orthogonal to it. The components orthogonal to  $\ket{\phi_A}$ can be expanded as a power series in time $t$. Since $\ket{\psi_z(0)}=\ket{\phi_A}$, the expansion to order $t^2$ can be written as
\begin{equation}
    \label{eq:psi_z_state_expansion}
    \ket{\psi_z(t)} = f_z(t)\ket{\phi_A} + t \ket{\eta_{z}} +t^2 \ket{\eta'_{z}} + \bigO{t^3},
\end{equation}
where $\braket{\phi_A|\eta_z}=\braket{\phi_A|\eta'_z}=0$ and $f_z(t)$ accounts for the time dependence of the component along $\ket{\phi_A}$. We fix the local phase of each normalized projected state by requiring $\braket{\phi_A|\psi_z(t)}=f_z(t)\in\mathbb{R}_{\geq0}$. This is allowed since the choice of local phases does not affect $\rho_\mathcal{E}^{(k)}$. Imposing the normalization condition
\begin{equation}
    \label{eq:psi_z_normalization}
    \braket{\psi_z|\psi_z} = f_z(t)^2 +  \norm{\eta_z}^2 t^2 +\bigO{t^3} = 1
\end{equation}
yields
\begin{equation}
    f_z(t) = \parens{1 -\norm{\eta_z}^2 t^2 +\bigO{t^3}}^\frac{1}{2} =1 - \frac{1}{2}\norm{\eta_z}^2 t^2+\bigO{t^3},
\end{equation}
where $\norm{\eta_z}^2\equiv\braket{\eta_z|\eta_z}$. 

The overlap between the projected states is
\begin{align}
    \braket{\psi_{z_1}|\psi_{z_2}} &= f_{z_1}(t)f_{z_2}(t) + \braket{\eta_{z_1}|\eta_{z_2}}t^2 + \bigO{t^3} \nonumber\\
    &=1-\frac{1}{2}\Bigl(\norm{\eta_{z_1}}^2+\norm{\eta_{z_2}}^2 \nonumber\\
    &\hspace{3.5em}-2\braket{\eta_{z_1}|\eta_{z_2}}\Bigr)t^2+\bigO{t^3},
\end{align}
yielding
\begin{align}
    \label{eq:psi_z_overlap_normalized_squared}
    \abs{\braket{\psi_{z_1}|\psi_{z_2}}}^2
    &=1-\Bigl(\norm{\eta_{z_1}}^2+\norm{\eta_{z_2}}^2 \nonumber\\
    &\hspace{3.5em}-2\mathrm{Re}\braket{\eta_{z_1}|\eta_{z_2}}\Bigr)t^2+\bigO{t^3}\nonumber\\
    &=1-\norm{\eta_{z_1}-\eta_{z_2}}^2t^2+\bigO{t^3}.
\end{align}
Comparison with Eq.~\eqref{eq:delta} gives
\begin{equation}
    \label{eq:delta_eta}
    \Delta_{z_1,z_2}(t) =\norm{\eta_{z_1}-\eta_{z_2}}^2t^2+\bigO{t^3} = \bigO{t^2}.
\end{equation}
In general, we have $M_l = \bigO{t^{2l}}$.
This scaling can be used in Eq.~\eqref{eq:fk_f1_general} to obtain the main result of this paper.
\begin{result}[Entanglement-randomness relation]
    \label{res:1}
    Let $\mathcal{E}$ be the projected ensemble on $A$ generated by evolving the initial product state $\ket{\Psi_0} = \ket{\phi_A} \otimes \ket{\phi_B}$ for time $t$. Assuming $\ket{\phi_B}$ is supported on all the measurement basis states of the bath,
    \begin{align}
    F^{(k)}(t) &= 1-k\parens{1-F^{(1)}(t)}+\bigO{t^4} \label{eq:fk_f1_time_series}\\ 
    &=\parens{F^{(1)}(t)}^k+\bigO{t^4}\label{eq:fk_f1_time_series_1}.
    \end{align}
\end{result}
The second equality follows from the fact that $M_1(t) = \bigO{t^2}$ for initial product states, thus $\parens{F^{(1)}(t)}^k = (1 - M_1(t))^k = 1 - k M_1 {(t)} + \bigO{t^4}$ using Eq.~\eqref{eq:delta_1}, for fixed $k$. Note that Jensen's inequality implies $F^{(k)}(t)\geq\parens{F^{(1)}(t)}^k$, so the $\bigO{t^4}$ correction is non-negative. Since $F^{(1)}(t)$ is simply the subsystem purity of $A$, the early-time growth of quantum randomness in the projected ensemble is entirely determined by the growth of entanglement between subsystems $A$ and $B$ up to cubic order.

Let us write $H = H_A + H_B + H_{I}$, where $H_A$ ($H_B$) is the part of the Hamiltonian that acts non-trivially on subsystem $A$ ($B$), and $H_I$ is the interaction Hamiltonian between the two subsystems. The early-time behavior of the subsystem purity $F^{(1)}(t)$ is~\cite{unanyan2010short,yang2018entanglement},
\begin{equation}
    F^{(1)}(t) = 1 - 2 \parens{\frac{t}{T_{\text{ent}}}}^2 + \bigO{t^3},
\end{equation}
where $T_{\text{ent}}$ is the \textit{entanglement timescale}~\cite{yang2018entanglement} given by
\begin{equation}
    T_{\text{ent}} \equiv \norm{\Pi_0 H \ket{\Psi_0}}^{-1} = \norm{\Pi_0 H_I \ket{\Psi_0}}^{-1},
    \label{eq:Tent}
\end{equation}
with $\Pi_0 \equiv (I_A-\ket{\phi_A}\bra{\phi_A}) \otimes (I_B-\ket{\phi_B}\bra{\phi_B})$ the tensor product of projectors onto the orthogonal subspaces of the initial states. Note that we have written $T_{\text{ent}}$ in a different, but equivalent, form compared to Ref.~\cite{yang2018entanglement}. Intuitively, $T_\text{ent}^{-1}$ measures the rate at which the interaction Hamiltonian jointly drives both subsystems out of their respective initial states, thereby generating bipartite entanglement. In fact, $T_{\text{ent}}$ also characterizes the early-time growth of entanglement as measured by \textit{any} quantum R\'{e}nyi entropy (for R\'{e}nyi indices $\alpha \geq 2$) of the subsystem density matrix $\rho_A$~\cite{cresswell2018universal}.

Substituting this expansion into Eq.~\eqref{eq:fk_f1_time_series} expresses the initial growth of randomness in terms of the entanglement timescale.

\begin{result}[Entanglement timescale for randomness growth]
    \label{res:2}
    Let $\mathcal{E}$ be the projected ensemble on $A$ generated by evolving the initial product state $\ket{\Psi_0} = \ket{\phi_A} \otimes \ket{\phi_B}$ for time $t$. Assuming $\ket{\phi_B}$ is supported on all the measurement basis states of the bath,
    \begin{align}
        \label{eq:framepot_series}
        F^{(k)}(t)&=1 -2k \parens{\frac{t}{T_{\text{ent}}}}^2 +\bigO{t^3}
    \end{align}
\end{result}
Unlike Result~\ref{res:1}, this explicit expansion has an $\bigO{t^3}$ rather than an $\bigO{t^4}$ remainder. The cubic contribution to the subsystem purity can be computed but lacks a simple physical interpretation. The moment order $k$ therefore rescales only the leading decay, while the governing timescale remains $T_{\text{ent}}$.

\subsection{Role of measurement basis}
Under the full-support assumption, the above results show that the effect of the measurement basis on the early-time growth of quantum randomness can first appear at $\bigO{t^4}$. While we do not quantify how the measurement basis affects the $\bigO{t^4}$ term, we provide some intuition for how the measurement basis affects the validity of the short-time approximation of $F^{(k)}(t)$. To derive Result~\ref{res:1}, we assumed that the initial state $\ket{\phi_B}$ has full support in the measurement basis, which we will now revisit. The unnormalized projected state has the Taylor expansion
\begin{equation}
    \ket{\tilde{\psi}_z(t)} = \braket{z|\phi_B}\ket{\phi_A} - i t \ket{\chi_{z}} + \bigO{\norm{H}^2 t^2},
\end{equation}
where $\ket{\chi_z} = (I_A \otimes \bra{z}_B) H \ket{\Psi_0}$. The relative perturbation of the second term to the first is bounded by
\begin{equation}
    \frac{\norm{t\ket{\chi_z}}}{\sqrt{p(z,0)}} \leq \frac{t\norm{H}}{\sqrt{p(z,0)}}.
\end{equation}
Let $p_{\min} = \min_z p(z,0)$ be the minimum probability of measuring any outcome $z$ at time $t=0$. For the relative perturbation to be small, we require $t\norm{H}\ll\sqrt{p_{\min}}$. Note that this is only a sufficient condition, and the actual time window of validity may be larger.

\subsection{Numerical results for mixed-field 1D Ising model}\label{sec:MFIM}

We test Eq.~\eqref{eq:fk_f1_time_series} numerically using the one-dimensional mixed-field Ising Hamiltonian (MFIM) with open boundary conditions:
\begin{equation}
    \label{eq:mfim-hamiltonian}
    H = \sum_{i=1}^{N_A + N_B}\parens{ h_y Y_i + h_z Z_i} + J\sum_{i=1}^{N_A+N_B-1}Y_iY_{i+1}.
\end{equation}
Here, $A$ contains sites $1,\ldots,N_A$, while $B$ contains sites $N_A+1,\ldots,N_A+N_B$. The operators $Y_i$ and $Z_i$ are Pauli operators acting on qubit $i$. We choose $\parens{h_y,h_z,J}=\parens{0.8090,0.9045,1}$, which lies in the nonintegrable, chaotic regime~\cite{kim2013ballistic}.

Figure~\ref{fig:mfim_result} compares $F^{(k)}$ with $(F^{(1)})^k$ for $k=2,3$, using MFIM dynamics and the initial state
\begin{equation}
    \label{eq:allplusstate}
    \ket{\Psi_0}={\ket{+}}^{\otimes N_A} \otimes {\ket{+}}^{\otimes N_B} = \ket{{+}}^{\otimes (N_A + N_B)},
\end{equation}
Result~\ref{res:2} gives $F^{(k)}(t)=1-2kJ^2t^2+\bigO{t^3}$, independent of $N_A$ and $N_B$. The numerical results agree with the approximation $F^{(k)}(t)\approx\parens{F^{(1)}(t)}^k$ at early times, up to approximately $Jt=0.2$.
Additional comparisons across $N_A$ and initial states on subsystem $A$, including longer-time dynamics, are shown in Appendix~\ref{app:numerics}.

The initial state in Eq.~\eqref{eq:allplusstate} satisfies both Eqs.~\eqref{eq:Psi_0} and \eqref{eq:psi_0_assumption}. By contrast, the initial state
\begin{equation}
    \label{eq:all0state}
    \ket{\Psi_0} = \ket{+}^{\otimes N_A} \otimes \ket{0}^{\otimes N_B},
\end{equation}
is still a product state but places all $N_B$ bath qubits in a single computational-basis state, violating the full-support assumption in Eq.~\eqref{eq:psi_0_assumption}. To examine this breakdown numerically, we interpolate between the two limits using amplitudes that define an exponentially biased computational-basis distribution parameterized by $\gamma\in\mathbb{R}_{\geq0}$:
\begin{equation}
    \label{eq:boltzmannstates1}
    \ket{\Psi_0, \gamma} \equiv \ket{+}^{\otimes N_A} \otimes \frac{1}{\sqrt{\mathcal{N}_\gamma}}\sum_{z=0}^{2^{N_B}-1} e^{-\gamma z}\ket{z}.
\end{equation}
Here, 
\begin{equation}
    \mathcal{N}_\gamma =  \sum_{z=0}^{2^{N_B}-1} e^{-2\gamma z}
\end{equation}
is the normalization factor. Here we hold $\ket{\phi_A}=\ket{+}^{\otimes N_A}$ fixed and vary only the bath state:
\begin{align}
    \ket{\Psi_0, \gamma=0} &= \ket{+}^{\otimes N_A} \otimes {\ket{+}}^{\otimes N_B},\\
    \ket{\Psi_0, \gamma\rightarrow \infty} &= \ket{+}^{\otimes N_A} \otimes \ket{0}^{\otimes N_B}.
\end{align}
For finite $\gamma$ the state has formal full support, but $p_{\min}=e^{-2\gamma(2^{N_B}-1)}/\mathcal{N}_\gamma$ becomes exponentially small. %

\begin{figure}[t]
    
    \centering
    \includegraphics[width=\columnwidth]{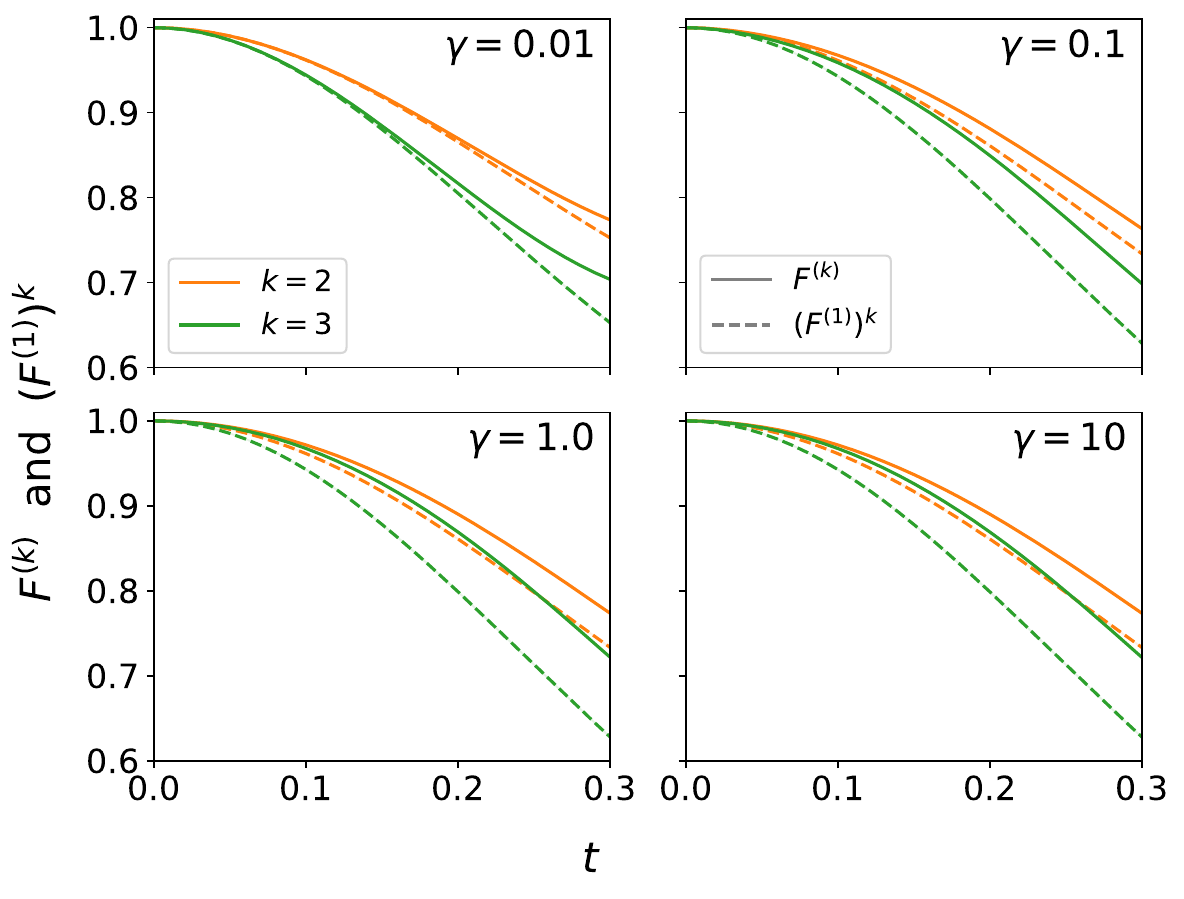}
    \caption{Numerical validation of Eq.~\eqref{eq:fk_f1_time_series_1} for initial states $\ket{\Psi_0, \gamma}$ of subsystem sizes $N_A=5$ and $N_B=6$ evolved under the 1D mixed-field Ising model. Plots of $F^{(k)}(t)$ (solid lines) and $(F^{(1)}(t))^k$ (dashed lines) for different values of $\gamma$ are shown. Higher values of $\gamma$ shrink $p_{\min}=e^{-2\gamma(2^{N_B}-1)}/\mathcal{N}_\gamma$, thus reducing the short-time window over which the approximation holds.
    }
    \label{fig:mfim_gamma_states}
\end{figure}

Figure~\ref{fig:mfim_gamma_states} shows that the finite-time discrepancy grows with $\gamma$, consistent with the shrinking $p_{\min}$ and the resulting reduction of the useful short-time window. The full-support assumption holds for every finite $\gamma$ and fails only in the limit $\gamma\to\infty$. %

\section{Unitarily invariant ensembles and spectral variance}\label{sec:unitarily_invariant}

The preceding results hold for arbitrary Hamiltonians. We now specialize to a unitarily invariant ensemble $\U$, whose distribution is invariant under $H\mapsto UHU^\dag$ for any unitary $U$. We assume that $\Eset{H\sim\U}\norm{H}^4<\infty$, for the results below to hold. A prominent example is the Gaussian Unitary Ensemble (GUE), an ensemble of random Hermitian matrices with complex Gaussian entries~\cite{mehta2004random}. The GUE is a standard model of quantum chaos and exhibits level repulsion and the dip--ramp--plateau structure of the spectral form factor~\cite{cotler2017chaos}. Previous work showed that projected ensembles generated by GUE dynamics form $k$-designs at certain $\bigO{1}$ and late times~\cite{ghosh2025design}; see Ghosh et al.~\cite{ghosh2025design} for details, including a generalization to $\U$. We instead study the early-time behavior for a general unitarily invariant ensemble, thereby complementing their analysis.

\begin{figure*}[t]
    \centering
    \includegraphics[scale =0.47]{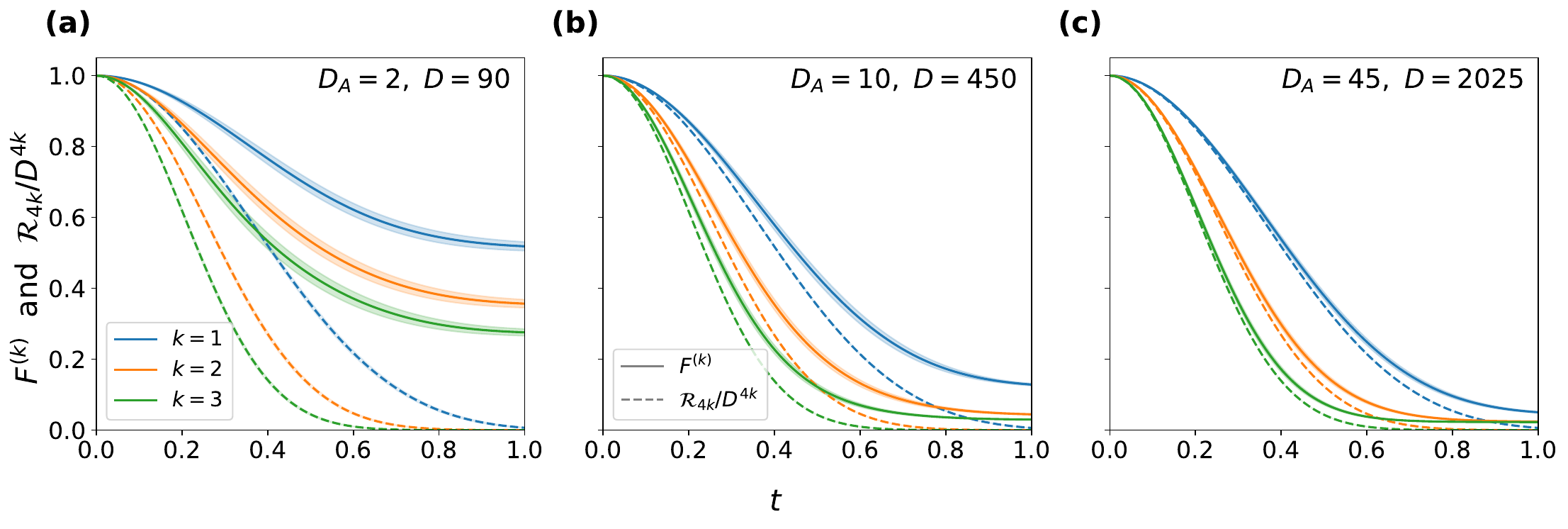}
    \caption{Ensemble-averaged $F^{(k)}(t)$ (solid lines) and $\sff{4k}(t)/D^{4k}$ (dashed lines) for 10 GUE samples, with $D_B=45$ and varying $D_A$. Error bands depict one standard deviation over GUE samples. Better finite-dimension agreement is observed for larger $D_A$.}
    \label{fig:gue_result}
\end{figure*}

A general Hamiltonian $H \in \U$ can be diagonalized as $H= U \Lambda U^\dagger$, where $U$ is a Haar random unitary, and $\Lambda$ is the diagonal matrix containing the eigenvalues of $H$. Moreover, $U$ and $\Lambda$ are statistically independent. To compute the average $k$th frame potential over the ensemble $\U$, we first perform the average over $U$ and then over $\Lambda$. The average over $U$ can be performed using Weingarten calculus~\cite{collins2022weingarten}, allowing us to relate $\E_{H \sim \U} F^{(k)}$ to the spectral properties of $H$.

The ensemble-averaged frame potential is also an even function of time. To see this, choose bases in which the initial state and the bath measurement projectors are invariant under complex conjugation. For each $H$, complex conjugation gives $F_H^{(k)}(t)=F_{H^*}^{(k)}(-t)$. Since $H$ and $H^*$ have the same distribution for a unitarily invariant ensemble, all odd powers of $t$ vanish after ensemble averaging.

\begin{result}[Randomness growth for unitarily invariant ensembles]

    \label{res:3}
    Let $\mathcal{E}$ be the projected ensemble on $A$ generated by evolving the initial product state $\ket{\Psi_0}=\ket{\phi_A}\otimes\ket{\phi_B}$ for time $t$ under a Hamiltonian $H$ drawn from a unitarily invariant ensemble $\U$. Assuming $\ket{\phi_B}$ is supported on all the measurement basis states of the bath, the ensemble-averaged $k$th frame potential is

    \begin{equation}
        \label{eq:gue_frame_pot}
        \Eset{H\sim \U} F^{(k)}(t) = 1 - \frac{2k(D_A-1)(D_B-1)}{D(D^2-1)} \Sigma^2 t^2 +\bigO{t^4}.
    \end{equation}

    Here, 
    \begin{equation}
        \Sigma^2\equiv \mathbb{E}_{\Lambda}\parens{D \Tr{\Lambda^2}-\parens{\Tr{\Lambda}}^2 }, 
    \end{equation}
    is $D^2$ times the ensemble-averaged variance of the eigenvalues of $H$. %
\end{result}
Details of the derivation are provided in Appendix~\ref{app:res3}.

\subsection{Relation to spectral form factors}
For a Hamiltonian $H\in\U$ with eigenvalues $\{\lambda_i\}_{i=1}^D$, define the analytically continued infinite-temperature partition function $Z(t)\equiv\Tr(e^{-iHt})=\sum_i e^{-i\lambda_i t}$. Following the standard convention, the $2k$-point spectral form factor is~\cite{cotler2017chaos}
\begin{equation}
    \label{eq:sff_Z}
    \sff{2k}(t) = \braket{\abs{Z(t)}^{2k}}_{\U} =\braket{\parens{Z(t) Z^*(t)}^{k}}_{\U}.
\end{equation}
Here, the average is taken over $\U$. The $2k$-point spectral form factor probes spectral correlations up to order $2k$. We use Eq.~\eqref{eq:sff_Z} to calculate $\sff{2k}$. Expanding $e^{-iHt}$ in time and evaluating the average of each term yields the following early-time relation between the ensemble-averaged $F^{(k)}$ and $\sff{4k}$ (see Appendix~\ref{app:res4} for details).

\begin{result}[Randomness growth and spectral form factor]
    \label{res:4}
    Let $\mathcal{E}$ be the projected ensemble on $A$ generated by evolving the initial product state $\ket{\Psi_0}=\ket{\phi_A}\otimes\ket{\phi_B}$ for time $t$ under a Hamiltonian $H$ drawn from a unitarily invariant ensemble $\U$. Assuming $\ket{\phi_B}$ is supported on all the measurement basis states of the bath, the ensemble-averaged $k$th frame potential satisfies
    \begin{equation}
        \label{eq:gue_frame_pot_sff}
        \Eset{H\sim \U} F^{(k)}(t) = \frac{\sff{4k}(t)}{D^{4k}}+\frac{2k \Sigma^2 t^2}{D^2}\parens{1-q_{A,B}}+\bigO{t^4},
    \end{equation}
    where $D=D_A D_B$ and $q_{A,B}\equiv D(D_A-1)(D_B-1)/(D^2-1)$.
    For $D_A, D_B \gg 1$, the second term in Eq.~\eqref{eq:gue_frame_pot_sff} can be neglected, giving
    \begin{equation}
        \label{eq:gue_frame_pot_sff_approx}
        \Eset{H\sim \U} F^{(k)}(t) \approx \frac{\sff{4k}(t)}{D^{4k}}+\bigO{t^4}
    \end{equation}
    to leading order in the joint short-time and large-dimension limit.
\end{result}
\vspace*{1em}
This result directly connects the early-time growth of randomness in projected ensembles with that of the global unitary dynamics. For any ensemble of unitaries $\mathcal{U}$, the frame potential $\mathcal{F}^{(k)}$ quantifies its randomness relative to the Haar ensemble~\cite{roberts2017chaos}:
\begin{equation}
    \norm{\Eset{U \sim \mathcal{U}}{U^{\otimes k} \otimes U^{\dag\otimes k}} - \Eset{U \sim \text{Haar}}{U^{\otimes k} \otimes U^{\dag\otimes k}}}_2^2 = \mathcal{F}^{(k)} - k!,
\end{equation}
where the frame potential for unitaries is defined as
\begin{equation}
    \mathcal{F}^{(k)} \equiv \Eset{U,V \sim \mathcal{U}}{\abs{\Tr{U^\dag V}}^{2k}}.
    \label{eq:unitary_frame_pot}
\end{equation}
Here, we assume that $k \leq D$ such that $\mathcal{F}^{(k)}_\text{Haar} = k!$. This is analogous to Eq.~\eqref{eq:HSdistance}, but now applied to unitaries. Here, we let $\mathcal{U}$ be the ensemble of unitaries generated by the Hamiltonians in $\U$, i.e., $\mathcal{U} = \{e^{-iHt} : H \in \U\}$, for a fixed time $t$. At early times, the $4k$-point spectral form factor and the $2k$th unitary frame potential satisfy
\begin{equation}
    \label{eq:sff_unitary_frame_pot}
    \frac{\sff{4k}(t)}{D^{4k}}
    =\frac{\sqrt{\mathcal{F}^{(2k)}(t)}}{D^{2k}}+\bigO{t^4}.
\end{equation}
We can therefore rewrite Eq.~\eqref{eq:gue_frame_pot_sff_approx} (for large $D_A, D_B$) as
\begin{equation}
    \Eset{H\sim \U} F^{(k)}(t) \approx \frac{\sqrt{\mathcal{F}^{(2k)}(t)}}{D^{2k}}+\bigO{t^4}.
\end{equation}
This equation expresses a ``$2k \to k$'' correspondence: the early-time growth of randomness in the projected ensemble at order $k$ is controlled by the global unitary ensemble at order $2k$.

This correspondence is complementary, but not equivalent, to previous ``$2k \to k$'' theorems~\cite{ghosh2025design,mok2026nature}. Those results state that the projected ensemble forms an approximate state $k$-design if the initial state is evolved under a global unitary drawn from an approximate unitary $2k$-design. In our case, the early-time unitary ensemble does not necessarily form an approximate $2k$-design.

Moreover, the previous theorems provide a sufficient (but not necessary) condition. Our result instead gives a direct quantitative relation between the randomness of the projected and global unitary ensembles, albeit only at early times and for Hamiltonians drawn from $\U$.

\subsection{Numerical results for GUE}
Figure~\ref{fig:gue_result} compares $\Eset{H\sim\gue}F^{(k)}$ with $D^{-4k}\sff{4k}$ for $k=1,2,3$. We fix $D_B=45$ and vary $D_A\in\bparens{2,10,45}$, finding better agreement as $D_A$ increases. This agreement requires $D_A,D_B\gg1$, but not $D_B\gg D_A$; the latter condition is needed only for the formation of approximate projected $k$-designs.

\begin{figure}[H]
    \centering
    \includegraphics[width=\columnwidth]{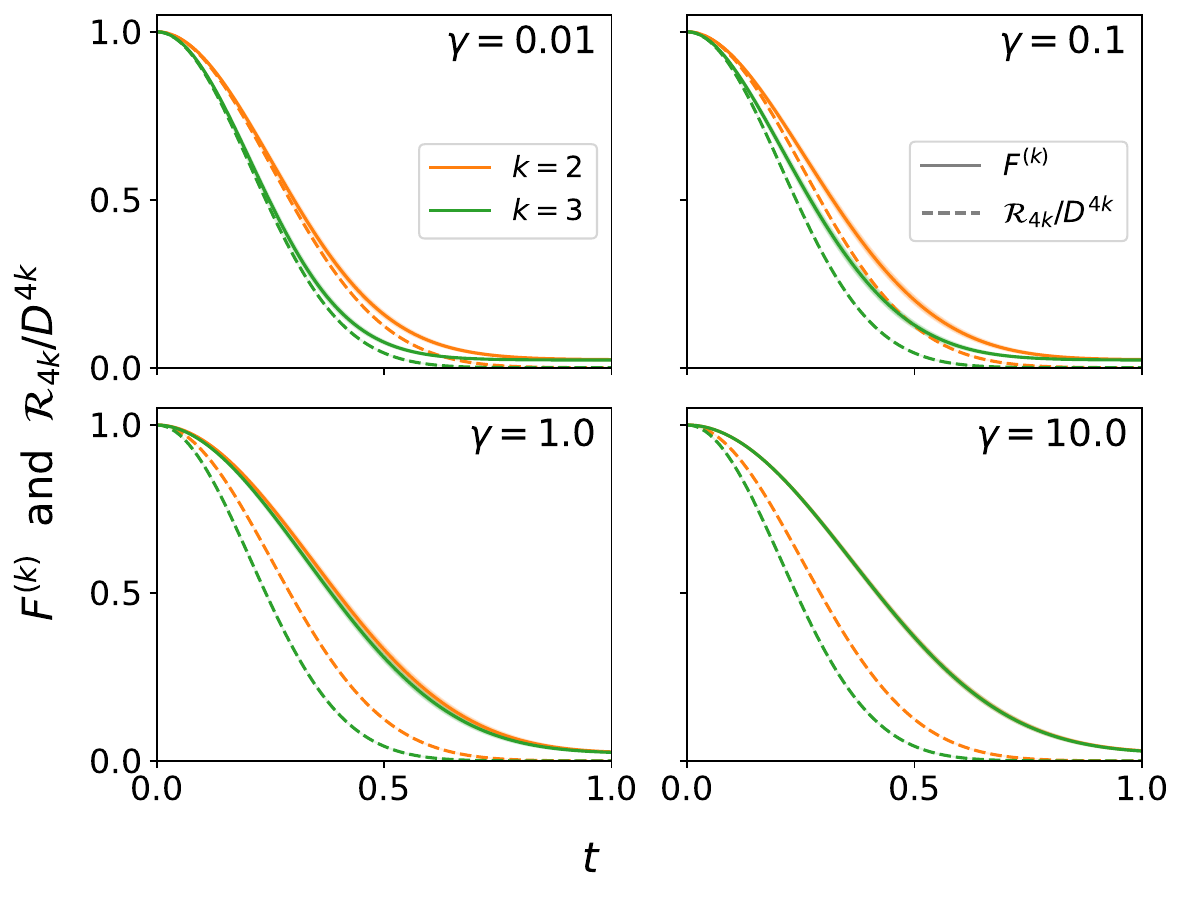}
    \caption{Numerical validation of the theory for initial states $\ket{\Psi_0, \gamma}$ evolved under GUE Hamiltonians. Results are shown for $D_A=D_B=45$. The early-time values of $F^{(k)}(t)$ (solid lines) and $\mathcal{R}_{4k}(t)/D^{4k}$ (dashed lines) begin to diverge at larger values of $\gamma$.
    }
    \label{fig:gue_gamma_states}
\end{figure}

As discussed in Sec.~\ref{sec:general}, the analytical results assume full support in the measurement basis [Eq.~\eqref{eq:psi_0_assumption}]. Figure~\ref{fig:gue_gamma_states} compares $\Eset{H\sim\gue}F^{(k)}$ with $D^{-4k}\sff{4k}$ for the initial states $\ket{\Psi_0,\gamma}$ defined analogously to Eq.~\eqref{eq:boltzmannstates1},
\begin{equation}
    \label{eq:boltzmannstates}
    \ket{\Psi_0, \gamma} \equiv \ket{+}_{D_A} \otimes \frac{1}{\sqrt{\mathcal{N}_\gamma}}\sum_{z=0}^{D_B-1} e^{-\gamma z}\ket{z}.
\end{equation}
Here, $\mathcal{N}_\gamma$ is the normalization factor and $\ket{+}_{D_A}=D_A^{-1/2}\sum_{x=0}^{D_A-1}\ket{x}$. As in Fig.~\ref{fig:mfim_gamma_states}, the finite-time agreement worsens as $\gamma$ increases and $p_{\min}$ decreases. Every finite $\gamma$ retains formal full support, but its short-time window of validity shrinks for larger $\gamma$; the assumption fails only in the limit $\gamma\to\infty$.
The corresponding approximation times are shown in Fig.~\ref{fig:app_approximation_time} of Appendix~\ref{app:numerics}.

\section{Conclusion}\label{sec:conclusion}
Most studies of quantum randomness in projected ensembles have focused on late-time equilibrium behavior. Here, we have shown that, provided the initial bath state has full support in the measurement basis, entanglement generated between the subsystems entirely controls the early-time growth of randomness up to cubic order. Because entanglement is basis independent, measurement-basis effects can first enter at fourth order. This behavior contrasts sharply with the late-time regime, where the measurement basis plays a crucial role in determining whether the projected ensemble forms an approximate $k$-design~\cite{liu2025coherence,mok2026nature}. Although we do not quantify the basis dependence at quartic and higher orders, our numerics show improved finite-time agreement for initial states with more uniform support in the measurement basis.

For Hamiltonians drawn from a unitarily invariant ensemble, such as the GUE, we additionally relate the projected-ensemble frame potential at order $k$ to the $4k$-point spectral form factor, equivalently the $2k$th frame potential of the global unitary ensemble. This ``$2k \to k$'' correspondence complements previous work on the regime in which the projected ensemble forms an approximate $k$-design. Whether the correspondence extends to generic chaotic Hamiltonians remains open.

Natural directions for future work include extending the expansion to higher orders, where measurement-basis dependence can first appear under the full-support assumption, and generalizing beyond the infinite-temperature setting to Hamiltonians with symmetries and conservation laws, for which the equilibrium projected ensemble is expected to approach a (generalized) Scrooge ensemble rather than the Haar ensemble.

\bibliography{bib}

\appendix
\clearpage      
\onecolumngrid

\section{Explicit expression for the early-time frame potential: proof of Result~\ref{res:2}}\label{app:res2}

\subsection{The entanglement timescale $T_{\text{ent}}$}

Any Hamiltonian $H$ can be decomposed into tensor products of Hermitian operators $A_n$ and $B_n$ that act only on subsystems $A$ and $B$ respectively:
\begin{equation}
    H =  \sum_n A_n \otimes B_n.
    \label{eq:app_H_decomp}
\end{equation}
Ref.~\cite{yang2018entanglement} defines the entanglement timescale $T_{\text{ent}}$ given by
\begin{equation}
    T_{\text{ent}}^{-2} = \sum_{m,n} \parens{\braket{A_m A_n}-\braket{A_m}\braket{A_n}}\parens{\braket{B_m B_n}-\braket{B_m}\braket{B_n}},
\end{equation}
where the expectation values are computed for a product initial state $\ket{\Psi_0}=\ket{\phi_A}\otimes\ket{\phi_B}$. We will show that this expression matches our definition of $T_{\text{ent}}$ given in Eq.~\eqref{eq:Tent}. First, note that,
\begin{align}
    \braket{\Psi_0|H \Pi_0 H |\Psi_0}
    &= \braket{\Psi_0|H \parens{I_A - \ket{\phi_A}\bra{\phi_A}}\otimes\parens{I_B - \ket{\phi_B}\bra{\phi_B}} H |\Psi_0}\nonumber\\
    &=\braket{H^2} -\braket{\Psi_0|H\parens{I_A\otimes \ket{\phi_B}\bra{\phi_B}}H|\Psi_0}-\braket{\Psi_0|H\parens{\ket{\phi_A}\bra{\phi_A}\otimes I_B}H|\Psi_0} +  \braket{H}^2 .
\end{align}
Each term can be computed in terms of the decomposition given in Eq.~\eqref{eq:app_H_decomp}.
\begin{align}
    \braket{H} &= \sum_n \braket{\phi_A \otimes\phi_B|A_n \otimes B_n|\phi_A\otimes\phi_B}\nonumber\\
    &=\sum_n \braket{\phi_A|A_n|\phi_A}\braket{\phi_B|B_n|\phi_B}\nonumber\\
    &= \sum_n \braket{A_n}\braket{B_n},
\end{align}
and so,
\begin{equation}
    \braket{H}^2 =  \sum_{m,n} \braket{A_n}\braket{B_n}\braket{A_m}\braket{B_m}.
\end{equation}
Similarly,
\begin{equation}
    \braket{H^2} = \braket{\sum_{m,n} \parens{A_m A_n}\otimes\parens{B_m B_n}} = \sum_{m,n} \braket{A_m A_n}\braket{B_m B_n}.
\end{equation}
The cross terms are:
\begin{align}
    \braket{\Psi_0|H\parens{I_A\otimes \ket{\phi_B}\bra{\phi_B}}H|\Psi_0}
    &= \sum_{m,n} \parens{\bra{\phi_A}\otimes\bra{\phi_B}}\parens{A_m \otimes B_m}\parens{I_A\otimes \ket{\phi_B}\bra{\phi_B}}\parens{A_n \otimes B_n}\parens{\ket{\phi_A}\otimes\ket{\phi_B}}\nonumber\\
    &= \sum_{m,n}\braket{\phi_A|A_m A_n|\phi_A}\braket{\phi_B|B_m|\phi_B}\braket{\phi_B|B_n|\phi_B}\nonumber\\
    &= \sum_{m,n}\braket{A_m A_n} \braket{B_m}\braket{B_n},
\end{align}
and, similarly,
\begin{equation}
    \braket{\Psi_0|H\parens{\ket{\phi_A}\bra{\phi_A}\otimes I_B}H|\Psi_0} = \sum_{m,n}\braket{A_m}\braket{A_n}\braket{B_m B_n}.
\end{equation}

Putting these together, we obtain
\begin{align}
    \braket{\Psi_0|H \Pi_0 H |\Psi_0} &= \sum_{m,n} \braket{A_m A_n}\braket{B_m B_n}-\braket{A_m A_n} \braket{B_m}\braket{B_n}-\braket{A_m}\braket{A_n}\braket{B_m B_n}+\braket{A_n}\braket{B_n}\braket{A_m}\braket{B_m}\nonumber\\
    &=  \sum_{m,n} \parens{\braket{A_m A_n}-\braket{A_m}\braket{A_n}}\parens{\braket{B_m B_n}-\braket{B_m}\braket{B_n}}\nonumber\\
    &= T_{\text{ent}}^{-2} .
\end{align}
Note that $\Pi_0$ is an orthogonal projector since it is a tensor product of $I_A-\ket{\phi_A}\bra{\phi_A}$ and $I_B-\ket{\phi_B}\bra{\phi_B}$, which are projectors onto the orthogonal complements of $\ket{\phi_A}$ and $\ket{\phi_B}$, respectively. Thus, $\Pi_0^\dagger\Pi_0=\Pi_0^2 =\Pi_0$, and therefore,
\begin{equation}
    T_{\text{ent}} = \braket{\Psi_0|H \Pi_0\Pi_0 H |\Psi_0}^{-\frac12} = \norm{\Pi_0 H \ket{\Psi_0}}^{-1},
\end{equation}
as claimed. 

As noted in Ref.~\cite{yang2018entanglement}, terms with $A_n =I_A$ (or $B_n = I_B$) do not contribute to $T_{\text{ent}}^{-2}$ since,
\begin{equation}
    \braket{A_m I_A}-\braket{A_m}\braket{I_A} = 0
\end{equation}
(and similarly for $B$). We can collect such terms and write
\begin{equation}
    H = H_A \otimes I_B + I_A\otimes H_B + H_I
\end{equation}
as defined in the main section. This means that $H_A$ and $H_B$ do not contribute to $T_{\text{ent}}^{-2}$, and therefore,
\begin{equation}
    T_{\text{ent}} = \norm{\Pi_0 H \ket{\Psi_0}}^{-1} = \norm{\Pi_0 H_I \ket{\Psi_0}}^{-1}.
\end{equation}
This can also be seen by explicitly showing that $ \Pi_0 H\ket{\Psi_0} = \Pi_0 H_I\ket{\Psi_0}$. 

Using our form of $T_{\text{ent}}$ in the time-series expansion derived in Ref.~\cite{yang2018entanglement}, the first frame potential for projected ensembles can be written as
\begin{equation}
    F^{(1)}(t) = 1 -2\parens{\frac{t}{T_{\text{ent}}}}^2 +\bigO{t^3} = 1 - 2 \braket{\Psi_0|H\Pi_0 H|\Psi_0}t^2 +\bigO{t^3}.
    \label{eq:app_F1_expansion}
\end{equation}

\subsection{Early-time behavior of $F^{(k)}(t)$ for $k>1$}

We will use Result~\ref{res:1} from the main text to provide an expression for $F^{(k)}(t)$ starting from Eq.~\eqref{eq:app_F1_expansion}. The proof of  Result~\ref{res:1} is outlined in Sec. \ref{sec:general} of the main text.
\begin{align}
    F^{(k)}(t) &= 1 - k(1- F^{(1)}(t))+ \bigO{t^4}\nonumber\\
    &=1- k \parens {2\parens{\frac{t}{T_{\text{ent}}}}^2 +\bigO{t^3}}+\bigO{t^4}\nonumber\\
    &= 1-2k\parens{\frac{t}{T_{\text{ent}}}}^2+\bigO{t^3}.
    \label{eq:app_F1_series}
\end{align}
This completes the proof of Result~\ref{res:2}.

\section{Average over unitarily invariant ensemble $\U$: proof of Result~\ref{res:3}}\label{app:res3}

In this section, we provide the required preliminaries and proof for deriving the ensemble average of the frame potential over any $\U$, starting from that of a generic Hamiltonian given in Eq.~\eqref{eq:framepot_series}.

\subsection{Preliminaries}

A unitarily invariant ensemble of $D$-dimensional Hamiltonians contains Hermitian matrices with a probability measure that is invariant under conjugation by any unitary matrix $U$ of the same dimension:
\begin{equation}
    P(H) =  P(U H U^\dagger) \qquad \forall\,\, U\in \mathcal{U}(D).
\end{equation}
A specific example is the Gaussian Unitary Ensemble, GUE($D,\mu,\sigma$), with $\mu\in\mathbb{R}$, $\sigma>0$, and probability density
\begin{equation}
    P(H)\propto \exp\left[-\frac{D}{2\sigma^2}\Tr{(H-\mu I)^2}\right].
\end{equation}
Equivalently, the diagonal entries are independent real Gaussian variables $H_{ii}\sim\mathcal{N}(\mu,\sigma^2/D)$, while for $i<j$ the off-diagonal entries are $H_{ij}=x_{ij}+iy_{ij}$, with $x_{ij}$ and $y_{ij}$ independent and distributed as $\mathcal{N}(0,\sigma^2/(2D))$, and $H_{ji}=H_{ij}^*$. This construction is invariant under unitary conjugation and satisfies $\mathbb{E}[H]=\mu I$.

$H\in\U$ is diagonalized by a unitary matrix $U$: $H=U\Lambda U^\dagger$. The diagonal matrix $\Lambda$ contains the eigenvalues of $H$ as its diagonal entries and the columns of $U$ are the corresponding eigenvectors. Now, consider some arbitrary unitary matrix $V$. Due to the unitary invariance of $\U$, we must have that,
\begin{equation}
    P(H) = P(U \Lambda U^\dagger) =  P(VU \Lambda U^\dagger V^\dagger) = P(VU \Lambda (VU)^\dagger)\qquad \forall \, V\in\mathcal{U}(D).
\end{equation}
Suppose we keep $\Lambda$ fixed. Then, the above equation implies that the diagonalizing unitaries $U$ of $\U$ are invariant under left multiplication by any fixed unitary $V$. Since $V^\dagger$ is also a unitary, invariance by right multiplication also holds. The unique distribution of unitary matrices that satisfies invariance under group multiplication on the left or right is the Haar distribution. Thus, the unitary matrices that diagonalize $H\in \U$ and encode the eigenvectors must be drawn from the Haar distribution. This is a common feature of any $\U$.

Moreover, note that the eigenvector distribution remains Haar conditioned on any $\Lambda$. Therefore, $\Lambda$ and $U$ must be statistically independent. The specific ensemble $\U$ is thus characterized by the eigenvalue distribution in $\Lambda$. For the GUE, the appropriately scaled eigenvalue density approaches Wigner's semicircle law at large $D$.

The statistical independence of $\Lambda$ and $U$ for a given $\U$ implies that ensemble averages factorize into an independent Haar average over the eigenvector matrices and an average over the eigenvalue matrices:
\begin{equation}
    \Eset{H\sim\U} f(H) = \Eset{\Lambda} \parens{\Eset{U\sim\text{Haar}}f(U\Lambda U^\dagger)}.
    \label{eq:app_factorize}
\end{equation}

\subsection{Weingarten calculus}

The Weingarten calculus is a useful tool for computing Haar averages of functions that are polynomial in the unitary matrices. The following formula, known as the $k$-fold twirl, is valid for an arbitrary operator $A$ acting on the $k$-fold Hilbert space $(\mathbb{C}^D)^{\otimes k}$:
\begin{equation}
    \Eset{U\sim\text{Haar}} \sparens{U^{\otimes k}A U^{\dagger \otimes k}} = \sum_{\sigma, \pi\in S_k} \text{Wg}(\sigma^{-1}\pi, D)\Tr{(A\hat{\pi}^\dagger)}\hat{\sigma}.
    \label{eq:app_twirl}
\end{equation}
Here, $S_k$ denotes the symmetric group of $k$ objects. The sum therefore runs over $(k!)^2$ ordered pairs of elements of $S_k$. For each $\tau\in S_k$, the $D^k \times D^k$ matrix $\hat{\tau}$ denotes its unitary representation on $(\mathbb{C}^D)^{\otimes k}$. The coefficients are given by the Weingarten functions $\text{Wg}(\tau, D)$, which are rational functions in $D$. This formula, along with some trace identities, allows one to compute the inner expectation in Eq.~\eqref{eq:app_factorize} for polynomial functions $f$.

\subsection{Averaging $F^{(k)}(t)$ over $\U$}

Let $\bparens{\ket{\eta}}$ denote an orthonormal basis in the full Hilbert space of dimension $D$. First, note that,
\begin{align}
  \braket{\Psi_0|H\Pi_0H|\Psi_0}
    &= \sum_{\eta}\braket{\Psi_0|H\Pi_0 \ket{\eta}\bra{\eta} H|\Psi_0} = \sum_{\eta} \braket{\Psi_0, \eta|H^{\otimes 2}\parens{\Pi_0 \otimes I}|\eta,\Psi_0}.
\end{align}
Here $\ket{\Psi_0, \eta} = \ket{\Psi_0}\otimes\ket{\eta}\in (\mathbb{C}^D)^{\otimes 2}$. We only need the expectation value of this matrix element to compute the average of frame potentials over $\U$ to $\bigO{t^2}$. By linearity of expectation, we have
\begin{align}
    \label{eq:app_matelem_exp}
    \Eset{H\sim\U} \braket{\Psi_0|H\Pi_0H|\Psi_0} = \sum_{\eta} \Eset{H\sim \U}  \braket{\Psi_0, \eta|H^{\otimes 2}\parens{\Pi_0 \otimes I}|\eta,\Psi_0} = \sum_{\eta}  \braket{\Psi_0, \eta|\Eset{H\sim \U} \parens{H^{\otimes 2}}\parens{\Pi_0 \otimes I}|\eta,\Psi_0}.
\end{align}

Computing $\Eset{H\in \U} \parens{H^{\otimes 2}}$ requires a double summation over $S_2$, whose elements are the identity permutation $e$ and the transposition $s$. Their unitary representations on $(\mathbb{C}^D)^{\otimes 2}$ are $\hat e=I_2$ and $\hat s=X_2$, respectively, where $X_2$ is the swap operator. These satisfy $X_2^\dagger=X_2^{-1}=X_2$ and $X_2^2=I_2$. The necessary Weingarten functions are
\begin{equation}
    \text{Wg}(1^2, D) = \frac{1}{D^2-1} \qquad \text{and} \qquad \text{Wg}(2, D) = -\frac{1}{D(D^2-1)}.
\end{equation}
Using Eq.~\eqref{eq:app_factorize} and Eq.~\eqref{eq:app_twirl}, we have that,
\begin{align}
    \Eset{H\in \U} \parens{H^{\otimes 2}}
    &= \Eset{\Lambda} \parens{\Eset{U\sim\text{Haar}} (U \Lambda U^\dagger)^{\otimes 2}}\nonumber\\
    &= \Eset{\Lambda} \parens{\Eset{U\sim\text{Haar}} \parens{U^{\otimes 2} \Lambda^{\otimes 2} U^{\dagger \otimes 2}}}\nonumber\\
    &= \Eset{\Lambda} \sum_{\sigma, \pi\in S_2} \text{Wg}(\sigma^{-1}\pi, D)\Tr{(\Lambda^{\otimes 2}\hat{\pi}^\dagger)}\hat{\sigma}\nonumber\\
    &= \Eset{\Lambda} \biggl(
    \text{Wg}(1^2, D) \Tr{\parens{\Lambda^{\otimes 2}}}I_2
    +\text{Wg}(2, D) \Tr{\parens{\Lambda^{\otimes 2}}}X_2\nonumber\\
    &\qquad +\text{Wg}(2, D) \Tr{\parens{\Lambda^{\otimes 2}X_2}}I_2
    +\text{Wg}(1^2, D) \Tr{\parens{\Lambda^{\otimes 2}X_2}}X_2
    \biggr)\nonumber\\
    &=\frac{1}{D^2-1}\Eset{\Lambda}\sparens{\parens{\parens{\Tr{\Lambda}}^2-\frac{1}{D}\Tr{\Lambda^2}}I_2+ \parens{-\frac{1}{D}\parens{\Tr{\Lambda}}^2+\Tr{\Lambda^2}}X_2}\nonumber\\
    &\equiv K_I I_2+K_X X_2,
\end{align}
where we have defined the scalars
\begin{equation}
    K_I \equiv \frac{1}{D^2-1}\Eset{\Lambda}\parens{\parens{\Tr{\Lambda}}^2-\frac{1}{D}\Tr{\Lambda^2}}
    \quad 
    \text{and}
    \quad
    K_X \equiv \frac{1}{D^2-1}\Eset{\Lambda}\parens{-\frac{1}{D}\parens{\Tr{\Lambda}}^2+\Tr{\Lambda^2}}.
\end{equation}

Substituting this into Eq.~\eqref{eq:app_matelem_exp} and noting that $X_2\ket{\Psi_0,\eta}=\ket{\eta, \Psi_0}$, we obtain
\begin{align}
    \Eset{H\sim\U} \braket{\Psi_0|H\Pi_0H|\Psi_0} 
    &= 
    K_I \sum_{\eta}  \braket{\Psi_0, \eta|I_2\parens{\Pi_0 \otimes I}|\eta,\Psi_0}
    +K_X \sum_{\eta}  \braket{\Psi_0, \eta|X_2\parens{\Pi_0 \otimes I}|\eta,\Psi_0}\nonumber\\
    &= K_I \sum_{\eta} \braket{\Psi_0|\Pi_0|\eta}\braket{\eta|\Psi_0} +  
    K_X \sum_{\eta} \braket{\Psi_0|\Psi_0} \braket{\eta|\Pi_0|\eta}\nonumber\\
    &=K_I \braket{\Psi_0|\Pi_0|\Psi_0} + K_X \Tr{\Pi_0}.
\end{align}
Since $(I_A - \ket{\phi_A}\bra{\phi_A})\ket{\phi_A}=0$, and similarly for $B$, $\Pi_0\ket{\Psi_0}=0$, so the first term in the sum vanishes. Noting that $\Pi_0$ is a projector, we obtain $\Tr{\Pi_0}=\Tr{\parens{I_A-\ket{\phi_A}\bra{\phi_A}}}\Tr{\parens{I_B-\ket{\phi_B}\bra{\phi_B}}}=(D_A-1) (D_B-1)$. Thus,
\begin{equation}
    \Eset{H\sim\U} \braket{\Psi_0|H\Pi_0H|\Psi_0} = \frac{(D_A-1)(D_B-1)}{D(D^2-1)}\Eset{\Lambda}\parens{D\Tr \Lambda^2-\parens{\Tr \Lambda}^2} \equiv \frac{(D_A-1)(D_B-1)}{D(D^2-1)}\Sigma^2.
\end{equation}
Finally, the ensemble average is even in time because complex conjugation maps $(H,t)$ to $(H^*,-t)$, while $H$ and $H^*$ have the same distribution in a unitarily invariant ensemble. Putting these results together, we obtain
\begin{equation}
    \Eset{H\sim\U} F^{(k)}(t) = 1 -2kt^2\Eset{H\sim\U} \braket{\Psi_0|H\Pi_0H|\Psi_0} + \bigO{t^4} = 1-\frac{2k(D_A-1)(D_B-1)}{D(D^2-1)}\Sigma^2t^2 +\bigO{t^4},
    \label{eq:app_Fk_avg}
\end{equation}
as claimed in Result~\ref{res:3}.

\section{Spectral form factor connection: proof of Result~\ref{res:4}}\label{app:res4}

\subsection{Taylor expansion of the $4k$-point spectral form factor}\label{app:r4_sff_expansion}

The ensemble average of the $4k$-point form factor can be computed to $\bigO{t^2}$ starting from Eq.~\eqref{eq:sff_Z}. The absolute square of the infinite-temperature partition function can be expressed to $\bigO{t^2}$ as follows:
\begin{align}
    \abs{Z(t)}^2 &= Z(t) Z^*(t)
    \nonumber\\
    &= \Tr{\parens{e^{iHt}}}\Tr{\parens{e^{-iHt}}}
    \nonumber\\
    &= \Tr\parens{I + iHt - \frac12 H^2 t^2 +\bigO{t^3}}\Tr\parens{I - iHt - \frac12 H^2 t^2+\bigO{t^3}}
    \nonumber\\
    &= \parens{D+it \Tr H -\frac{1}{2} t^2 \Tr{H^2}+\bigO{t^3}}
    \parens{D-it \Tr H -\frac{1}{2} t^2 \Tr{H^2}+\bigO{t^3}}\nonumber\\
    &=D^2 + t^2\parens{\parens{\Tr H}^2-D \Tr H^2}+\bigO{t^4}\nonumber\\
    &=D^2\parens{1 - \frac{1}{D^2}\Sigma_H^2 t^2+\bigO{t^4}}.
    \label{eq:app_sff_series}
\end{align}
Here, we have defined
\begin{equation}
    \Sigma_H^2 \equiv  D \Tr(H^2) -\parens{\Tr H}^2,
\end{equation}
which is $D^2$ times the variance of the eigenvalues of $H$. The $4k$-point spectral form factor of a single Hamiltonian is thus
\begin{align}
    R_{4k} \equiv \abs{Z(t)}^{4k} = \parens{\abs{Z(t)}^2}^{2k} =D^{4k}\parens{1 -\frac{2k}{D^2}\Sigma_H^2 t^2+\bigO{t^4}}.
\end{align}
The last equality follows from binomially expanding Eq.~\eqref{eq:app_sff_series}.

\subsection{Averaging the spectral form factor over $\U$}

Due to invariance of the trace under a basis change, we have
\begin{equation}
    \Sigma_H^2 =  D \Tr(H^2) -\parens{\Tr H}^2=D \Tr(\Lambda^2) -\parens{\Tr \Lambda}^2.
\end{equation}
Thus,
\begin{equation}
    \Eset{H\sim\U} \Sigma_H^2= \Eset{U\sim\text{Haar}}\parens{\Eset{\Lambda}\parens{D \Tr(\Lambda^2) -\parens{\Tr \Lambda}^2}} = \Sigma^2,
    \label{eq:app_Sigma_squared}
\end{equation}
and,
\begin{align}
    \frac{\mathcal{R}_{4k}}{D^{4k}}\equiv  \Eset{H\sim\U}\frac{R_{4k}}{D^{4k}}=1 -\frac{2k}{D^2}t^2\Eset{H\sim\U} \Sigma_H^2+\bigO{t^4}= 1- \frac{2k}{D^2}\Sigma^2 t^2+\bigO{t^4}.
    \label{eq:app_sff}
\end{align}

\subsection{\texorpdfstring{Relation to the unitary frame potential}{Relation to the unitary frame potential}}

The early-time frame potential for an ensemble of unitaries generated by a unitarily invariant ensemble $\U$ can be related to the ensemble-averaged spectral form factor. We first replace the variables $U,V$ in Eq.~\eqref{eq:unitary_frame_pot} with unitaries generated by their Hamiltonians:
\begin{equation}
    U = e^{-iH_1 t} \qquad\text{and}\qquad V=e^{-iH_2 t}.
\end{equation}
Thus, 
\begin{align}
    \mathcal{F}^{(k)}
    = \Eset{U,V\sim \mathcal{U}}\abs{\Tr U^\dagger V}^{2k}
    = \Eset{H_1, H_2 \sim \U} \abs{\Tr \parens{e^{iH_1 t}e^{-iH_2 t}}}^{2k}.
\end{align}
Using the Baker--Campbell--Hausdorff (BCH) formula, we have
\begin{align}
    e^{iH_1 t}e^{-iH_2 t}
    &=\exp\sparens{it(H_1 -H_2)+\frac{t^2}{2} [H_1, H_2 ]+\bigO{t^3}}\nonumber\\
    &= I + it(H_1 - H_2)+\frac{t^2}{2} [H_1, H_2 ] +\frac{1}{2!}\parens{it\parens{H_1-H_2}}^2+\bigO{t^3}\nonumber\\
    &= I + it(H_1-H_2)- \frac{t^2}{2}\parens{H_1^2+H_2^2-2H_1 H_2}+\bigO{t^3}.
\end{align}
Taking the trace, we obtain,
\begin{align}
    \Tr\parens{e^{iH_1 t}e^{-iH_2 t}}
    = D + it \parens{\Tr{H_1} -\Tr{H_2}}- \frac{t^2}{2}\parens{\Tr (H_1^2)+\Tr(H_2^2)-2\Tr(H_1 H_2)}+\bigO{t^3}.
\end{align}
Note that the odd powers of $t$ will contain purely imaginary terms due to $H$ being Hermitian. Thus, the modulus-squared of the above quantity will only contain even powers of $t$. 
\begin{align}
    \abs{\Tr\parens{e^{iH_1 t}e^{-iH_2 t}}}^{2}
    &= D^2 +t^2 \parens{\Tr H_1 -\Tr H_2}^2- D t^2 \parens{\Tr (H_1^2)+\Tr(H_2^2)-2\Tr(H_1 H_2)}
    +\bigO{t^4}
\end{align}
Upon raising to power $k$ using binomial expansion, we obtain,
\begin{align}
    \abs{\Tr\parens{e^{iH_1 t}e^{-iH_2 t}}}^{2k}
    &= D^{2k}\parens{1 + \frac{t^2}{D^2} \parens{\parens{\Tr H_1 -\Tr H_2}^2- D \parens{\Tr (H_1^2)+\Tr(H_2^2)-2\Tr(H_1 H_2)}
    }+\bigO{t^4}}^k\nonumber\\
    &= D^{2k}\parens{1 + \frac{kt^2}{D^2} \parens{\parens{\Tr H_1 -\Tr H_2}^2- D \parens{\Tr (H_1^2)+\Tr(H_2^2)-2\Tr(H_1 H_2)}
    }+\bigO{t^4}}\nonumber\\
    &= D^{2k}\parens{1- \frac{kt^2}{D^2}\parens{\Sigma_{H_1}^2 +\Sigma_{H_2}^2 - 2C_{H_1, H_2}}+\bigO{t^4} },
    \label{eq:app_unitary_frame_pot_expansion}
\end{align}
where
\begin{equation}
    C_{H_1, H_2} \equiv D \Tr(H_1 H_2)- (\Tr{H_1})(\Tr{H_2}),
\end{equation}
can be shown to be $D^2$ times the covariance between the eigenvalues $\lambda_{1,i}$ of $H_1$ and $\lambda_{2,j}$ of $H_2$ with a joint probability distribution defined by the overlap between their eigenvectors, $p(\lambda_{1,i}, \lambda_{2,j}) = \frac{\abs{\braket{1,i|2,j}}^2}{D}$.

Since $H_1$ and $H_2$ are independently and identically distributed in $\U$,  Eq.~\eqref{eq:app_Sigma_squared} gives us,
\begin{align}
    \Eset{H_1, H_2\sim\U} \Sigma_{H_i}^2 = \Eset{H_i\sim\U} \Sigma_{H_i}^2 = \Sigma^2,
\end{align}
for $i=1,2$. Using Weingarten calculus, we obtain,
\begin{align}
    \Eset{H\sim\U} H
    &= \Eset{\Lambda}\parens{\Eset{U\sim\text{Haar}}\parens{U\Lambda U^\dagger}} \nonumber\\
    &= \text{Wg}(1,D)\Eset{\Lambda}\Tr \parens{\Lambda} I\nonumber\\
     &= \frac1D\Eset{\Lambda}\Tr \parens{\Lambda} I.
    \label{eq:app_H_expectation}
\end{align}
Thus,
\begin{equation}
    \Eset{H_1, H_2\sim\U} H_1 H_2 = \parens{\Eset{H\sim \U}H}^2 = \frac{1}{D^2} \parens{\Eset{\Lambda} \Tr(\Lambda)}^2 I.
\end{equation}

These equations yield
\begin{align}
     \Eset{H_1, H_2\sim\U}C_{H_1, H_2} &\equiv D \Eset{H_1, H_2\sim\U} \Tr(H_1 H_2) - \Eset{H_1, H_2\sim\U}(\Tr{H_1})(\Tr{H_2}) \nonumber\\
     &= D\Tr\parens{\Eset{H_1, H_2\sim\U} H_1 H_2}- \parens{\Tr\parens{\Eset{H\sim\U} H}}^2 \nonumber\\
     &= D\frac{1}{D^2} \parens{\Eset{\Lambda} \Tr(\Lambda)}^2 \Tr(I)-\parens{\frac{1}{D}\Eset{\Lambda}\Tr(\Lambda)\Tr\parens{I}}^2 \nonumber\\
     &=\parens{\Eset{\Lambda}\Tr(\Lambda)}^2 - \parens{\Eset{\Lambda} \Tr(\Lambda)}^2\nonumber\\
     &=0.
\end{align}
Substituting all the computed expectation values in Eq.~\eqref{eq:app_unitary_frame_pot_expansion}, we obtain,
\begin{align}
    \frac{\mathcal{F}^{(k)}(t)}{D^{2k}}
    &= \Eset{H_1, H_2\sim \U} \frac{\abs{\Tr\parens{e^{iH_1 t}e^{-iH_2 t}}}^{2k}}{D^{2k}}\nonumber\\
    &= 1- \frac{kt^2} {D^2}\Eset{H_1, H_2\sim \U}\parens{\Sigma_{H_1}^2 +\Sigma_{H_2}^2 - 2C_{H_1, H_2}}+\bigO{t^4}\nonumber\\
    &= 1- \frac{2k\Sigma^2}{D^2} t^2 + \bigO{t^4}.
\end{align}
It follows that
\begin{equation}
    \frac{\mathcal{F}^{(2k)}(t)}{D^{4k}}
    =1-\frac{4k\Sigma^2}{D^2}t^2+\bigO{t^4}.
\end{equation}
Taking the square root and comparing with Eq.~\eqref{eq:app_sff} gives
\begin{equation}
    \frac{\sff{4k}(t)}{D^{4k}}
    =\frac{\sqrt{\mathcal{F}^{(2k)}(t)}}{D^{2k}}+\bigO{t^4},
\end{equation}
as stated in Eq.~\eqref{eq:sff_unitary_frame_pot}.

\subsection{Connection to $F^{(k)}(t)$}\label{app:r4_match}

Define
\begin{equation}
    q_{A,B}\equiv \frac{D(D_A-1)(D_B-1)}{D^2-1}.
\end{equation}
Then Eq.~\eqref{eq:app_Fk_avg} can be written exactly through quadratic order as
\begin{equation}
    \Eset{H\sim\U} F^{(k)}(t)
    =1-\frac{2k\Sigma^2}{D^2}q_{A,B}t^2+\bigO{t^4}.
\end{equation}
Combining this expression with Eq.~\eqref{eq:app_sff} gives
\begin{equation}
    \Eset{H\sim\U} F^{(k)}(t)
    =\frac{\sff{4k}(t)}{D^{4k}}
    +\frac{2k\Sigma^2 t^2}{D^2}\parens{1-q_{A,B}}+\bigO{t^4},
\end{equation}
as claimed in Result~\ref{res:4}. Assuming that the bath is larger, $D_A\leq D_B$, we have,
\begin{align}
    q_{A,B} = \frac{D D_A D_B}{D^2} \parens{1-\frac{1}{D_A}}\parens{1-\frac{1}{D_B}}\parens{1-\frac{1}{D^2}}^{-1} = 1 +\bigO{\frac{1}{D_A}}.
\end{align}

The $\bigO{t^2}$ correction between $\Eset{H\sim\U} F^{(k)}(t)$ and $\frac{\sff{4k}(t)}{D^{4k}}$ is $\frac{2k\Sigma^2t^2}{D^2}\bigO{\frac{1}{D_A}}$, which is suppressed in the limit $D_A, D_B \gg 1$.

\clearpage
\section{Additional numerical results} \label{app:numerics}

In this section, we present additional numerics to support our claims.
For all GUE simulations presented in this work, we use the centered ensemble, $\mu=0$, with the variance convention defined in Appendix~\ref{app:res3}.

\begin{figure}[H]
    \centering
    \includegraphics[width=0.8\columnwidth]{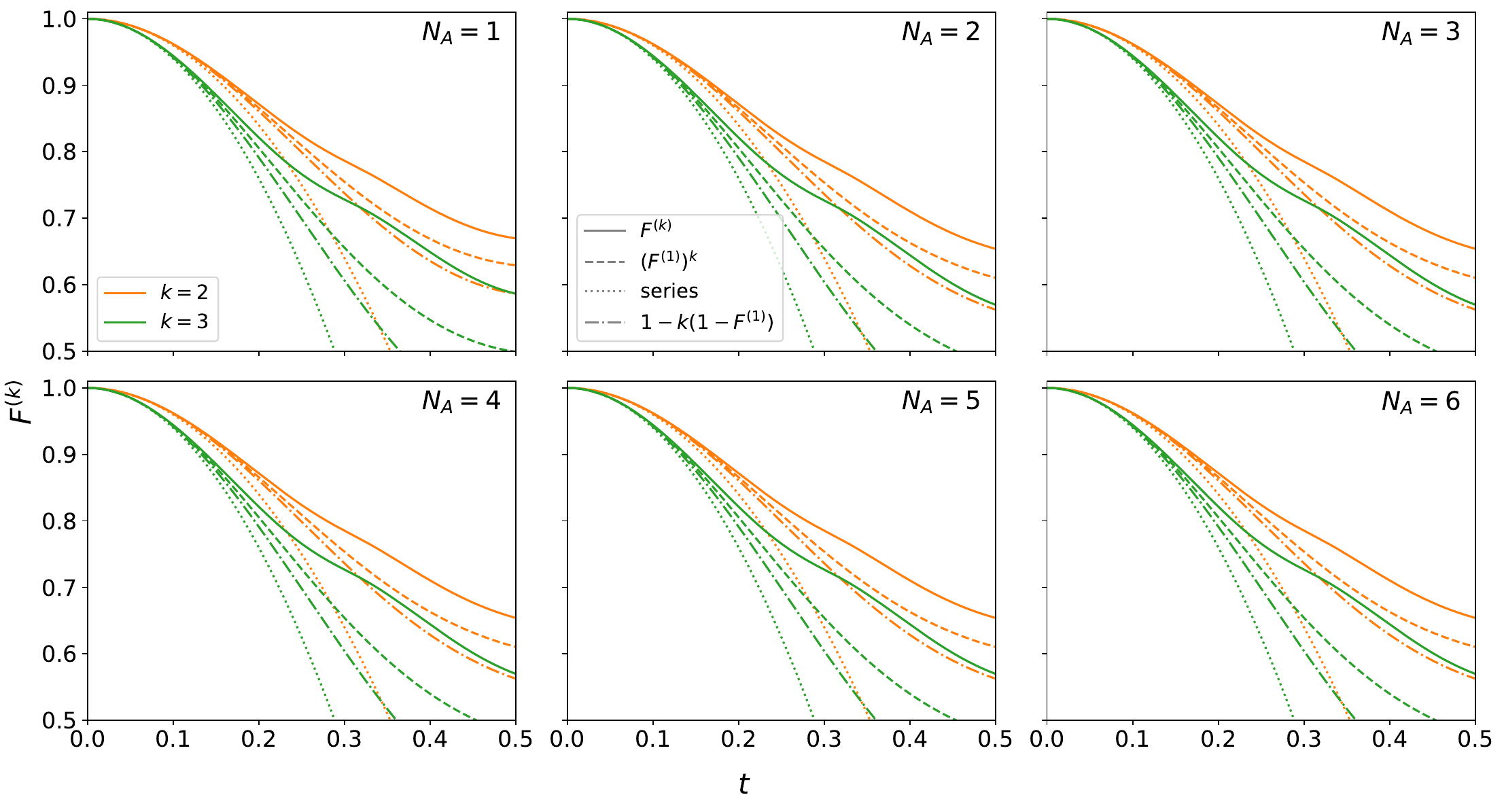}
    \caption{
    Early-time approximations to $F^{(k)}(t)$ for the 1D mixed-field Ising model with $k=2,3$, $N_B=6$, and $N_A=1,\ldots,6$. Solid lines show $F^{(k)}(t)$; dashed, dash-dotted, and dotted lines show $[F^{(1)}(t)]^k$, $1-k[1-F^{(1)}(t)]$, and $1-2k\braket{\Psi_0|H\Pi_0H|\Psi_0}t^2$, respectively. The initial state is $\ket{\Psi_0}=\ket{+}^{\otimes(N_A+N_B)}$. The approximations agree with $F^{(k)}(t)$ at early times.
    }
    \label{fig:app_na_varied}
\end{figure}
\begin{figure}[H]
    \centering
    \includegraphics[width=0.7\columnwidth]{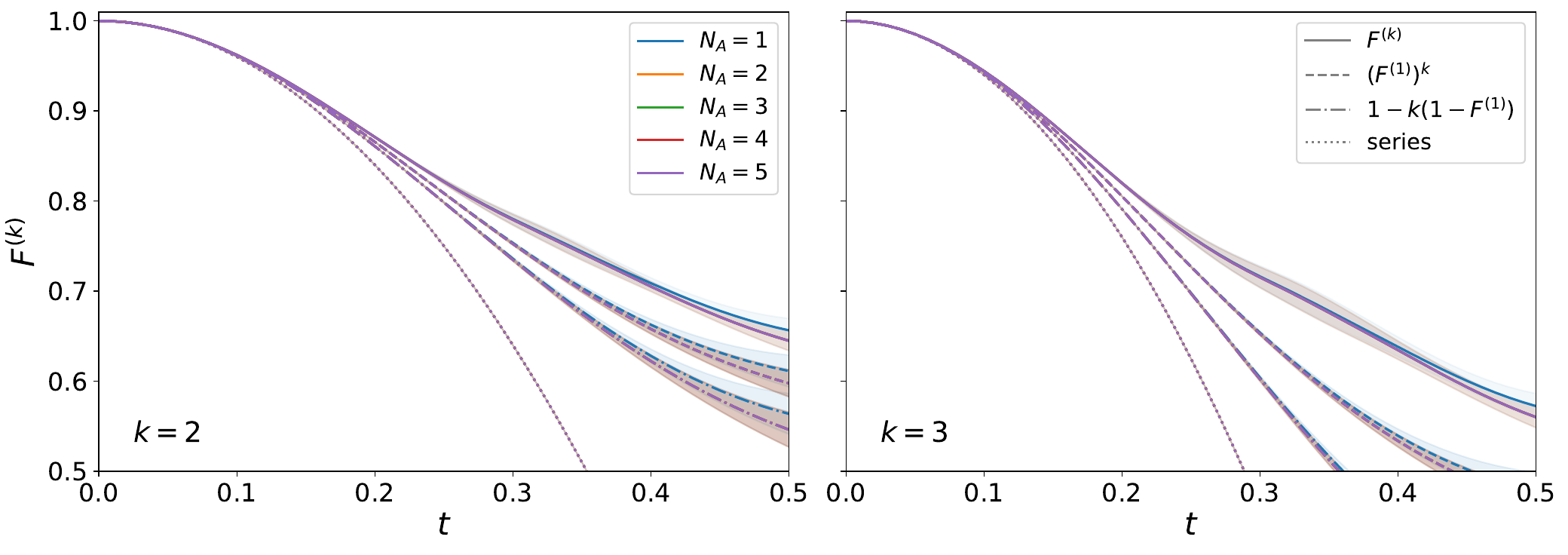}
    \caption{
    Early-time approximations from Fig.~\ref{fig:app_na_varied}, averaged over all $2^{N_A}$ computational-basis states $\ket{\phi_A}$, for (left) $k=2$ and (right) $k=3$. Shaded regions show one standard deviation over $\ket{\phi_A}$.
    }
    \label{fig:app_na_phia_varied}
\end{figure}
\begin{figure}[H]
    \centering
    \includegraphics[width=0.7\columnwidth]{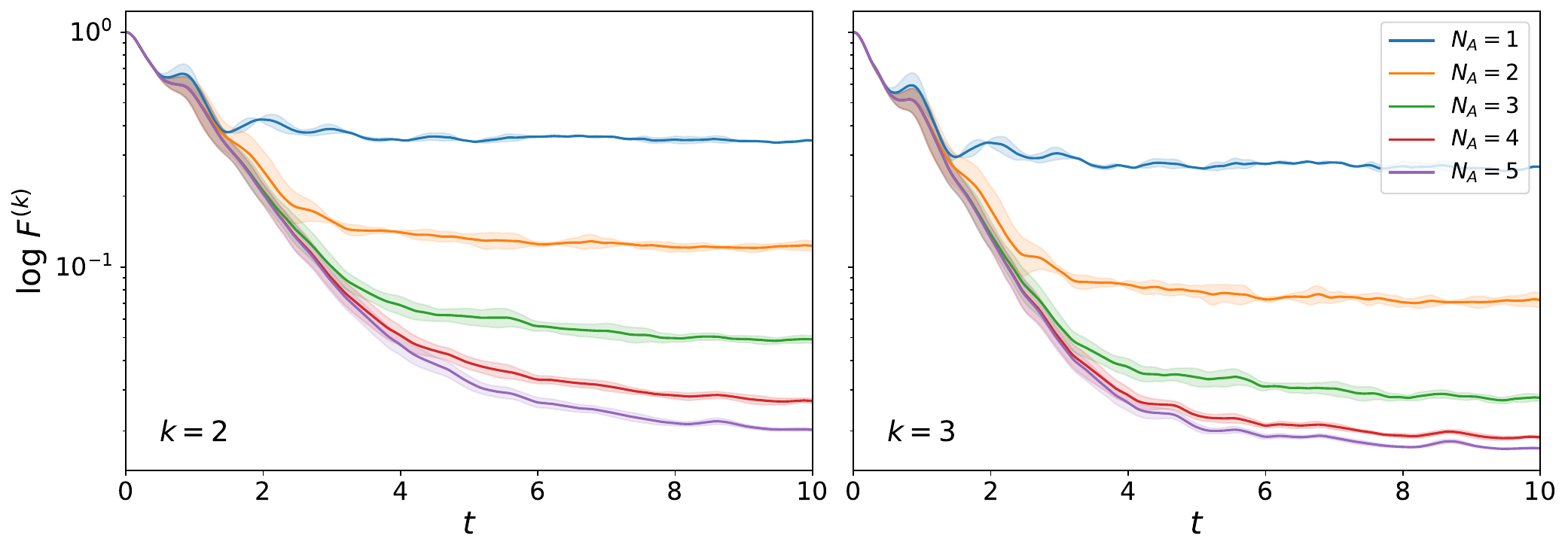}
    \caption{
    Frame potential $F^{(k)}(t)$ up to $Jt=10$ for the setup of Fig.~\ref{fig:app_na_phia_varied}, with (left) $k=2$ and (right) $k=3$. The shaded bands indicate one standard deviation over $\ket{\phi_A}$. Dependence on $N_A$ becomes visible at intermediate and late times but is negligible at the early-time scale shown in Fig.~\ref{fig:app_na_phia_varied}.
    }
    \label{fig:app_fk_long_time}
\end{figure}
\begin{figure}[H]
    \centering
    \includegraphics[width=0.7\columnwidth]{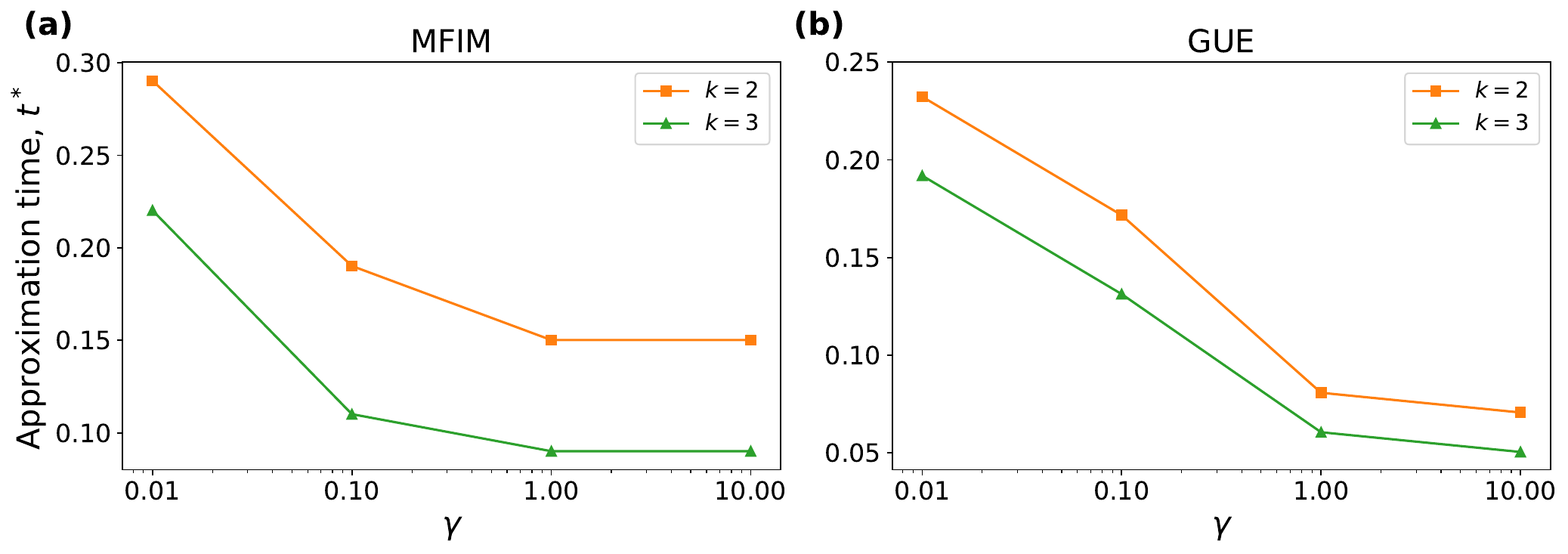}
    \caption{
    Approximation time $t^*$ versus $\gamma$ for $k=2,3$. We define $t^*$ as the largest sampled time such that the relative deviation from $F^{(k)}(t)$ remains below $2\%$ at all earlier sampled times. We compare $F^{(k)}(t)$ with $[F^{(1)}(t)]^k$ for the MFIM (left) and $\Eset{H\sim\gue}F^{(k)}(t)$ with $\sff{4k}(t)/D^{4k}$ for the GUE (right). In both cases, $t^*$ decreases or saturates as $\gamma$ increases.
    }
    \label{fig:app_approximation_time}
\end{figure}

\end{document}